\documentclass[11pt]{article}
\usepackage{amsmath}
\usepackage{amsfonts}

\begin{document}

\title{Dating a random walk: Statistics of the duration time of a random walk given its present position}

\author{Hern\'an Larralde \\ 
Instituto de Ciencias F\'isicas,UNAM. Apdo. Postal 48-3.\\
        C.P. 62251, Cuernavaca, Morelos, M\'exico.}
 
\date{\today}

%\pacs{05.40.-a}{Fluctuation phenomena, random processes, noise, and Brownian motion}
%\pacs{05.40.Fb}{Random walks and Levy flights}
%\pacs{05.60.-k}{Transport processes}
%\pacs{02.50.Ey}{Stochastic processes}
\maketitle

\begin{abstract}
  We consider the distribution of the duration time, the time elapsed
  since it began, of a diffusion process given its present position,
  under the assumption that the process began at the origin. For
  unbiased diffusion, the distribution does not exist (it is
  identically zero) for one and two dimensional systems. We find the
  explicit expression for the distribution for three and higher
  dimensions and discuss the behavior of the duration time statistics:
  we find that the expected duration time exists only for dimensions
  five and higher, whereas the variance becomes finite for seven
  dimensions and above. We then turn to the case of biased
  diffusion. The drift velocity introduces a new time scale and the
  resulting statistics arise from the interplay of the diffusive time
  scale and the drift time scale. For these systems all the moments
  exist and explicit expressions are presented and discussed for the
  expected duration time and its variance for all dimensions.
\end{abstract}

%=====================================================================
\section{Introduction}

The dating of events and objects, that is, the determination of the
time elapsed since an event occurred or the time since an object was
made (or last used), is fundamental in areas as diverse a geology and
archeology, to epidemiology and forensic science. A classic example is
Carbon dating, in which the ratio of $^{14}C$ to $^{12}C$ is used to
estimate the age of organic samples in archeology and geology
\cite{libby,bow}.  On the other hand, in forensic investigations, the
problem of estimating the time elapsed since the occurrence of an
event arises, for example, in the determination of the post mortem
interval (PMI), that is, the time since a person has died. The
estimation of the PMI is achieved by a variety of methods, ranging
from the temperature of the body, the study of the different stages
the body passes through after death -algor, livor and rigor mortis- to
the different stages of body decomposition. Other interesting forensic
techniques include using the development of flies and other insects,
as well as their larvae in the body to establish the
PMI\cite{foren}. Indeed, the guilt or innocence of a suspect, may
hinge on the accurate determination of the time of the events before,
during and after the crime.  In the context of engineering, the
determination of the time at which faults and errors occur in a
production process may help to pinpoint the causes behind each fault
and correct or prevent them from happening again in the
future\cite{shao}.  From the perspective of epidemiology, outbreaks
are detected when there are already several infected individuals. To
make things harder, in some cases, these individuals may have already
moved from the place were they caught the disease. In this situation,
knowing the time at which the epidemy began may be useful in
identifying the source of the disease \cite{CDC}. Other situations
related to the problem of dating the beginning of a diffusion process
would be the determination of when a leak occurred given the
dispersion of contaminant particles emanated from the leak or
establishing when a batch of marked or counterfeit bills were placed
in circulation.

The common feature of these examples is that given the present state
of a system -the ratio of radioactive isotopes in an archeological
sample, the number or age of fly larvae in a body, the number of
faulty products in a production line, the number and location of
infected individuales in an epidemic, the spatial distribution of
pollutants or of fake bills- one must infer when the evolution that
led to this state began. In all these cases, the amount of time
elapsed is determined through a precise knowledge of the initial
conditions and time evolution of the process in question. Of course,
since many random factors intervene in these systems, the inferred
time can, at best, be determined probabilistically. However, it
appears that not much attention has been devoted to the determination
of the statistics of the time elapsed given the present state of a
stochastic process; not even for one of the simplest and most widely
used of such processes: the random walk.  Random walks are, of course,
the paradigm of stochastic transport.  They have been used to model
the motion of living organisms, particles in suspensions, atoms and
molecules in and on solids, light in turbid media, among many other
things \cite{weiss}. Indeed, random walks represent the microscopic
process that gives rise to diffusive transport at macroscopic scales.
In this work we will be interested in the question of, given the
present position of one (or several) diffusing particles, when did the
process begin? That is, we want to find the probability distribution
$P(t|\bar{x},\bar{x_0})$ that the process has evolved for a time $t$,
given that it began at position $\bar{x_0}$ and is presently at
position $\bar{x}$. We will refer to this as the distribution of
``duration times''.

It should be noted that a process consisting of $M$ independent
$d$-dimensional diffusing particles is formally equivalent to the case
of a single particle diffusing in a $N=M\times d$ dimensional space.
Thus, in what follows, we will always refer to single diffusing
particles in $N$-dimensional spaces. Indeed, and perhaps not
unexpectedly, the properties of the duration time distribution will be
strongly dependent on the dimension of the space in which the walk
takes place, becoming more sharply peaked the higher the dimension. On
the other hand, we will also show that for unbiased diffusion, the
duration time distribution is not even defined (it is identically
zero) for 1 and 2-dimensional systems. This, we argue, is related to
Polya's recurrence theorem \cite{weiss, spitzer, feller}; namely that
unbiased random walks return to their initial possition with
probability 1 and thus return infinitely many times, in one and two
dimensional spaces, whereas their is a finite probability that the
walker will never return to its initial position in dimensions three
and higher.  We also show that the distribution of duration times for
unbiased diffusion exists but does not have a mean in 3 and 4
dimensions, whereas the variance of the distribution exists for
unbaised random walks in 7 and higher dimensions.  Next, we turn to
the distribution and the statistics of the duration time of biased
random walks. Since these are always transient, the distribution and
it's moments exist in all dimensions, we obtain their explicit expressions 
and discuss the limiting behaviors.

\section{Inferring the distribution of duration time for unbiased diffusion}

A central quantity in the theory of random walks is $P(\bar{x}|t,
\bar{x_0})$, the probability of finding a $N$-dimensional random
walker at position $\bar{x}$ given that it started at position
$\bar{x_0}$ and a time $t$ has elapsed \cite{weiss}. This probability
describes the transport properties of this process as a function of
time and has been extensively studied for over a century \cite{weiss,
  einst,strutt, chandra, kac, spitzer}. In what follows, the explicit
expressions for the distributions of duration times will be obtained
from Bayes' theorem \cite{jaynes} using $P(\bar{x}|t, \bar{x_0})$ and
an exponential prior duration time distribution
$p(t)=e^{-t/\lambda}/\lambda$. The choice of this prior distribution
-essentially the marginal distribution of duration times- is
arbitrary, but ensures that the duration time is not negative. Thus we
take the limit $\lambda \to\infty$ to have the most un-informative
distribution with positive domain. The idea behind this approach is,
first, to consider the duration time $t$ as a random variable. Then,
since we have interpreted $P(\bar{x}|t, \bar{x_0})$ as the conditional
position of the walker given that the duration time of the walk is
$t$, we multiply by the exponential prior distribution to construct a
joint distribution of elapsed times $t$ and positions. From this
joint distribution we can finally calculate the distribution of
elapsed times conditioned on the present position of the random walk
by dividing over the marginal distribution of the walker's
positions. However, the exponential prior distribution effectively
introduces a time scale $\lambda$ which is not intrinsic to the
diffusion process. While this choice may be realistic in certain
applications (for instance when the walkers are atoms which decay
radioactively, it is an exogenous time scale generated from additional
knowledge of the system). Thus by taking the limit $\lambda\to \infty$
we get rid of this extrinsic time scale and we are left with the
situation in which any (unconditioned) positive duration time is, in
some sense, equally likely. That is, the marginal time distribution,
the exponential, becomes vanishingly small, though still normalized,
ensuring only that the process began sometime in the past. Other
choices, for example taking $t$ uniformly distributed in the interval
$[0,T]$ and the taking the limit $T\to\infty$, give the same
results. Under these conditions then, the statistics of the duration
time of the random walk arise solely from the diffusion process.

For ease of calculation, we first consider the case of a N-dimensional
unbiased diffusion process, which corresponds to the continuum limit
of symmetric random walks with jump distributions with finite second
moment\cite{weiss, chandra, feller, risken, van}.  The probability
density for finding the walker at position ${\bar{x}}$ assuming it
began at the origin is given by:
\begin{equation}
P({\bar x}|t,0)=\frac{1}{(4\pi Dt)^{N/2}}e^{-\frac{1}{4Dt}\sum_{n=1}^{N} x_i^2}
\end{equation}
Thus, using an exponential prior, we have that the distribution of
duration times is \cite{grad}:
\begin{eqnarray}
P_\lambda(t|{\bar x},0)&=&\frac{e^{-\frac{1}{4Dt}\sum_{i=1}^{N} x_i^2}
e^{-t/\lambda}}{t^{N/2}\int_0^\infty e^{-\frac{1}{4D\tau}\sum_{i=1}^{N} x_i^2}
e^{-\tau/\lambda}\frac{d\tau}{\tau^{N/2}}}\\ &=&
\frac{e^{-\frac{1}{4Dt}\sum_{i=1}^{N} x_i^2} e^{-t/\lambda}}{2 t^{N/2}
\left(\frac{\lambda 
\sum_{i=1}^N x_i^2}{4D}\right)^{(1-N/2)/2} K_{1-N/2}
\left(\sqrt{\frac{\sum_{i=1}^N x_i^2}{D\lambda}}\right)}
\end{eqnarray}
where $K_\nu(z)$ denotes the modified Bessel function of order
$\nu$ \cite{abram}. In the limit $\lambda\to\infty$ we obtain:
\begin{equation}
P(t|{\bar x},0)=\frac{1}{t}\left(\frac{\sum_{i=1}^N 
x_i^2}{4Dt}\right)^{\frac{N}{2}-1} \frac{e^{-\frac{1}{4Dt}
\sum_{n=1}^{N} x_i^2}}{\Gamma(\frac{N}{2}-1)}, \qquad N>2
\end{equation}
while $P(t|{\bar x},0)\equiv 0$ for $N=1,2$. As mentioned above, the
reason the duration time distribution vanishes for one and two
dimensional random walks is because they are recurrent
\cite{weiss}. That is, with probability one, the walks return to the
vicinity of every point in the system, and thus, they return an
infinite number of times (though the average time between succesive
returns diverges). Therefore, by knowing the walker's position, there
is no way of knowing whether it has visited that point a finite or
infinite number of times in the past. Indeed, since it may have
visited its present position arbitrarily many times in the past,
essentially any arbitrarily large time may have elapsed since the walk
began and the distribution of duration times vanishes accordingly, so no
information on the duration of the process can be inferred from the
position of the walker in these cases. On the other hand, for three
and higher dimensional systems, there is a finite probability that the
walker never returns to the vicinity of a point, and if it does, it
returns in a finite mean time\cite{weiss}. Thus, the number of
previous visits to any given point is finite and the duration time for
the process has a vanishing probability of being indefinitely long.

When the duration time distribution is defined, it is straight forward
to see that the mean duration time does not exist for 3 and 4
dimensional systems; only becoming finite for $N\geq 5$:
\begin{equation}
<t|{\bar x}>=
\frac{2}{N-4}\left(\frac{\sum_{i=1}^N x_i^2}{4D}\right) \qquad N\geq 5
\end{equation}
The variance becomes finite for 7 and higher dimensional systems:
\begin{equation}
\frac{<t^2|{\bar x}>-<t|{\bar x}>^2}{<t|{\bar x}>^2}=\frac{2}{N-6}
\qquad N\geq 7
\end{equation}
so the higher the dimension of the system, the more precisely the time
elapsed since the beginning of the process can be determined or,
alternatively, the more accurately can the walker be used as a clock.

Another statistic that can be readily computed for systems of any
dimension $N$ is $t_{mpv}$, the most probable duration time, which is
given by
\begin{equation}
t_{mpv}({\bar x})= \frac{1}{N}\left(\frac{\sum_{i=1}^N
  x_i^2}{2D}\right)
\end{equation}

\section{The case of biased diffusion}

A more interesting case is that in which we attempt to date the
beginning of a biased diffusion process. In this case the process is
always transient and the distribution of duration times is defined in
every dimension. In contrast to the unbiased case, the biased case has
two distinct time scales (and hence infinitely many): one given by the
mean square displacement divided by the diffusion constant, another
given by the net displacement divided by the drift velocity. The
interplay of these scales will determine the statistics of the
duration time. Repeating the procedure described above, the
distribution of duration times given the position of the particle
diffusing in a N-dimensional system is given by:
\begin{equation}
P_\lambda(t|{\bar x},0)=\frac{e^{-\frac{1}{4Dt}\sum_{i=1}^{N}
    x_i^2-\frac{t}{4D}\sum_{i=1}^N v_i^2}}{2 t^{N/2}
  \left(\frac{\sum_{i=1}^N x_i^2}{\sum_{i=1}^N
    v_i^2}\right)^{(1-N/2)/2} K_{1-N/2}
  \left(\frac{1}{2D}\sqrt{\left(\sum_{i=1}^N x_i^2\right)
    \left(\sum_{i=1}^N v_i^2\right)}\right)}
\label{dist_v}
\end{equation}
The exponential factors induce cutoffs which insure that all the
moments of the distribution exist.  In particular, the mean duration
time can be shown to be
\begin{equation}
<t|{\bar x}>=\left(\frac{\sum_{i=1}^N x_i^2}{\sum_{i=1}^N v_i^2}\right)^{1/2}
\frac{K_{2-N/2}\left(\frac{1}{2D}\sqrt{\left(\sum_{i=1}^N x_i^2\right)
\left(\sum_{i=1}^N v_i^2\right)}\right)}{K_{1-N/2}
\left(\frac{1}{2D}\sqrt{\left(\sum_{i=1}^N x_i^2\right)
\left(\sum_{i=1}^N v_i^2\right)}\right)}
\end{equation}
and the variance is
\begin{eqnarray}
<t^2|{\bar x}> - <t|{\bar x}>^2 &=&
\left(\frac{\sum_{i=1}^N x_i^2}{\sum_{i=1}^N v_i^2}\right)\left[
\frac{K_{3-N/2}\left(\frac{1}{2D}\sqrt{\left(\sum_{i=1}^N x_i^2\right)
\left(\sum_{i=1}^N v_i^2\right)}\right)}{K_{1-N/2}
\left(\frac{1}{2D}\sqrt{\left(\sum_{i=1}^N x_i^2\right)
\left(\sum_{i=1}^N v_i^2\right)}\right)}\right. \nonumber \\ 
&&-
\left.
\left(\frac{K_{2-N/2}\left(\frac{1}{2D}\sqrt{\left(\sum_{i=1}^N x_i^2\right)
\left(\sum_{i=1}^N v_i^2\right)}\right)}{K_{1-N/2}
\left(\frac{1}{2D}\sqrt{\left(\sum_{i=1}^N x_i^2\right)
\left(\sum_{i=1}^N v_i^2\right)}\right)}\right)^2~\right]
\end{eqnarray}
which are exact, but admittedly not the most informative of formulae.
To have a better understanding of the behavior of these statistics, we
focus on the limits of large and small values of the arguments of the
Bessel functions.

For large values of the arguments of the Bessel functions, to leading
order, we have \cite{abram}
\begin{equation}
<t|{\bar x}>\sim \left(\frac{\sum_{i=1}^N x_i^2}{\sum_{i=1}^N v_i^2}\right)^{1/2}
~~~~ {\rm for}~~~~\left(\sum_{i=1}^N x_i^2\right)
\left(\sum_{i=1}^N v_i^2\right)>> 4D^2
\end{equation}
whereas in the same limit, the variance behaves as
\begin{eqnarray}
<t^2|{\bar x}> - <t|{\bar x}>^2 \sim 2D
\frac{\left(\sum_{i=1}^N x_i^2\right)^{1/2}}{\left(\sum_{i=1}^N 
v_i^2\right)^{3/2}}
\end{eqnarray}
so the relative fluctuations are small:
\begin{eqnarray}
\frac{<t^2|{\bar x}> - <t|{\bar x}>^2}{<t|{\bar x}>^2} \sim 
\frac{2D}{\left(\sum_{i=1}^N 
v_i^2\right)^{1/2}\left(\sum_{i=1}^N x_i^2\right)^{1/2}}<<1
\end{eqnarray}
Naively, this behavior can be understood by noting that for the
systems under consideration, very large displacements are mostly due
to the effect of the drift, the diffusive contribution playing an
increasingly negligible role, so the duration time can be accurately
estimated as the net displacement divided by the magnitud of the drift
velocity.

A richer variety of phenomena occur in the regime in which the
arguments of the Bessel functions are small. In this case both the
diffusion and the drift can contribute to the displacement to
different degrees depending on the dimension of the system. In this
regime, the mean duration time is given by
\begin{displaymath}
<t|{\bar x}>\sim
\left \{
\begin{array}{ll}
\frac{1}{2}\frac{4D}{\sum_{i=1}^N v_i^2} &\qquad\textrm{N=1}\\ \\
\frac{4D}{\left(\sum_{i=1}^N v_i^2\right) \ln\left[2D/
\left(\sum_{i=1}^N x_i^2\right)^{1/2} \left(\sum_{i=1}^N v_i^2\right)^{1/2}
\right]} &\qquad\textrm{N=2} \\ \\
\left(\frac{\sum_{i=1}^N x_i^2}{\sum_{i=1}^N v_i^2}\right)^{1/2} &
\qquad\textrm{N=3}\\ \\
\frac{\left(\sum_{i=1}^N x_i^2\right)}{4D} \ln\left[2D/
\left(\sum_{i=1}^N x_i^2\right)^{1/2} \left(\sum_{i=1}^N v_i^2\right)^{1/2}
\right] &\qquad \textrm{N=4}\\ \\
\frac{2}{N-4}\left(\frac{\sum_{i=1}^N x_i^2}{4D}\right) &
\qquad\textrm{N$\geq$ 5}
\end{array}
\right.
\end{displaymath}
Interestingly, for small enough displacements in one dimensional
systems, the expected duration time is independent of the
displacement, whereas for 5 or higher dimensional systems, the
duration time is independent of the drift velocity as long as
$\left(\sum_{i=1}^N x_i^2\right) \left(\sum_{i=1}^N
v_i^2\right)<<4D^2$.  In this same limit, the behavior of the variance
has to be evaluated case by case up to six dimensional systems; to
leading order we have:
\begin{displaymath}
<t^2|{\bar x}>-<t|{\bar x}>^2 \sim 
\left \{
\begin{array}{ll}
\frac{8 D^2}{\left(\sum_{i=1}^N v_i^2\right)^{2}} & \qquad\textrm{N=1}\\ \\
\frac{8D^2}{\left(\sum_{i=1}^N v_i^2\right)^2 \ln\left[2D/
\left(\sum_{i=1}^N x_i^2\right)^{1/2} \left(\sum_{i=1}^N v_i^2\right)^{1/2}
\right]} &\qquad\textrm{N=2} \\ \\
\frac{2D\left(\sum_{i=1}^N x_i^2\right)^{1/2}}{\left(
\sum_{i=1}^N v_i^2\right)^{3/2}} &\qquad\textrm{N=3}\\ \\
\frac{\sum_{i=1}^N x_i^2}{\sum_{i=1}^N v_i^2 } &\qquad\textrm{N=4} \\ \\ 
\frac{\left(\sum_{i=1}^N x_i^2\right)^{3/2}}{2D\left(
\sum_{i=1}^N v_i^2\right)^{1/2}} &
\qquad \textrm{N=5}\\ \\
\frac{\left(\sum_{i=1}^N x_i^2\right)^2}{8D^2}\ln\left[\frac{2D}{
\left(\sum_{i=1}^N x_i^2\right)^{1/2}\left(\sum_{i=1}^N v_i^2\right)^{1/2}}
\right] &\qquad\textrm{N=6}\\ \\
\frac{\left(\sum_{i=1}^N x_i^2\right)^2}{2D^2(N-4)^2(N-6)} &
\qquad\textrm{N$\geq$ 7}
\end{array} \right.
\end{displaymath}
On the other hand, the most probable duration time is given by the
expression:
\begin{eqnarray}
t_{mpv}({\bar x})=\frac{1}{\sum_{i=1}^N v_i^2}\left[\left( D^2N^2+
\left(\sum_{i=1}^N x_i^2\right)\left(\sum_{i=1}^N v_i^2\right)\right)^{1/2}
-DN\right]
\end{eqnarray}
This quantity ranges from
\begin{eqnarray}
t_{mpv}({\bar x})\sim\frac{\left(\sum_{i=1}^N x_i^2\right)}{2DN} \qquad {\rm for~~}
 D^2N^2 >> \left(\sum_{i=1}^N x_i^2\right)\left(\sum_{i=1}^N v_i^2\right)
\end{eqnarray}
to
\begin{eqnarray}
t_{mpv}({\bar x})\sim\frac{\left(\sum_{i=1}^N x_i^2\right)^{1/2}}{\left(\sum_{i=1}^N v_i^2\right)^{1/2}} \qquad {\rm for~~}
 D^2N^2 << \left(\sum_{i=1}^N x_i^2\right)\left(\sum_{i=1}^N v_i^2\right)
\end{eqnarray}
as expected. Note that this large displacement behavior coincides with
the mean value $<t|{\bar x}>$ in the same limit.

It is worth noting that the term in ${\bar x}\cdot{\bar v}$ cancels
out in the distribution of duration times Eq. (\ref{dist_v}), so that 
if the walker is biased in only one component, that particular
component plays no special r\^ole in determining the duration
time. Indeed, somewhat surprisingly, even though the drift breaks the
symmetry of the systems, the duration time distribution remains
rotationally invariant around the origin.

\section{Summary and Conclusions}

In summary, we have considered the question of the duration of a
diffusion process given its present state under the assumption that
the process began at the origin. This is tantamount to establishing
when the process began, given the present position of the diffusing
particle. We find that in the case of unbiased diffusion, the
distribution does not exist for one and two dimensional systems and
argued that this is due to the recurrent nature of unbiased random
walks in these dimensions. We find the explicit distribution for three
and higher dimensions and discuss the behavior of the duration time
statistics for these systems: since the distribution decays slowly, it
turns out that the expected duration time exists only for dimensions
five and higher, whereas the variance becomes finite for seven
dimensions and above. Not surprisingly, the system becomes a better
clock as the dimension increases.

Next we focused on the statistics of the duration time for a
N-dimensional biased diffusion process given the position of the
diffusing particle. The drift velocity in these systems introduces a
new time scale and the resulting statistics arise from the interplay
of the diffusive time scale and the drift time scale. For these
systems all the moments exist and explicit expressions are presented
for the expected duration time and its variance for all dimensions.

Future work will consider the statistics of duration time for process
that evolve by anomalous diffusion, due, say, to a power law
distribution of jump lengths in the case of superdiffusion, or to a
long tailed distribution of waiting times for subdiffusive systems
\cite{weiss}.  Other interesting and more complicated extensions could
involve considering the statistics of the duration time for diffusive
processes in other geometries; for example, by evaluating the effect
that a reflecting wall has on the distribution of duration times, or
by considering diffusion on a nontrivial substrate, like a fractal
\cite{havlin,jp}. Also, the dating of the beginning of other
stochastic processes not related to transport might be of interest. For
example, when did the ``first'' ancestor of a family line live, given
the present number of descendants.

\end{document}